\documentclass{article}
\usepackage{srcltx}
\usepackage{amsmath}
\usepackage{amssymb}
\usepackage{amsfonts}
\usepackage{latexsym}
\usepackage{textcomp}
\usepackage{appendix}
\usepackage{multirow}
\usepackage{booktabs}
\usepackage{url}
\usepackage{array}
\usepackage{marvosym}
\usepackage{graphics}
\usepackage{graphicx}
\title{The (in)visible hand in the Libor market: an Information Theory approach}

\author{Aurelio Fern\'andez Bariviera\\ \scriptsize{Department of Business, Universitat Rovira i Virgili, Av. Universitat 1, 43204 Reus, Spain} \\ \scriptsize{\ttfamily aurelio.fernandez@urv.net}   \and M. Bel\'en Guercio \\  \scriptsize{Instituto de Investigaciones Econ\'omicas y Sociales del Sur, UNS-CONICET.} \\  \scriptsize{12 de Octubre y San Juan, B8000CTX Bah\'{\i}a Blanca, Argentina.} \\ \scriptsize{Universidad Provincial del  Sudoeste (UPSO).} \\ \scriptsize{ Alvarado 328, B8000CJH Bah\'ia Blanca, Argentina} \and Lisana B. Martinez \\   \scriptsize{Instituto de Investigaciones Econ\'omicas y Sociales del Sur, UNS-CONICET.} \\  \scriptsize{12 de Octubre y San Juan, B8000CTX Bah\'{\i}a Blanca, Argentina.} \\ \scriptsize{Universidad Provincial del  Sudoeste (UPSO).} \\ \scriptsize{ Alvarado 328, B8000CJH Bah\'ia Blanca, Argentina} \and Osvaldo A. Rosso \\ \scriptsize{Instituto de F\'{\i}sica, Universidade Federal de Alagoas (UFAL). } \\ 
\scriptsize{BR 104 Norte km 97, 57072-970 Macei\'o, Alagoas, Brazil. }\\ \scriptsize{Instituto Tecnol\'ogico de Buenos Aires (ITBA),} \\ \scriptsize{Av. Eduardo Madero 399, C1106ACD Ciudad Aut\'onoma de Buenos Aires, Argentina.}}

\begin{document}
\maketitle

\begin{abstract}

This paper analyzes several interest rates time series from the United Kingdom during the period 1999 to 2014. The analysis is carried out using a pioneering statistical tool in the financial literature: the complexity-entropy causality plane. This representation is able to classify different stochastic and chaotic regimes in time series. We use sliding temporal windows to assess changes in the intrinsic stochastic dynamics of the time series. Anomalous behavior in the Libor is detected, especially around the time of the last financial crisis, that could be consistent with data manipulation. 

\end{abstract}%

\section{Introduction  \label{sec:intro}}

The London Interbank Offered Rate (Libor) was established in 1986 by the British Banking Association (BBA). Since then it has been a benchmark interest rate for pricing several derivative financial instruments that are traded on exchanges worldwide. In fact, Libor is one of the major reference interest rates not only for the interbank market but also for other transactions. Many financial contracts link the obligations of the contracting parties to Libor evolution.

On May 29, 2008, Mollenkamp and Whitehouse \cite{MollenkampWhitehouse} published an article on the Wall Street Journal casting doubts on the transparency of Libor settings and implying that published rates were lower than those reported by credit default swaps (CDS). Investigations by market authorities such as the US Department of Justice and the European Commission detected data manipulation and imposed severe fines to banks involved in such an illegal procedure.
The existing financial literature has provided no conclusive evidence of manipulation of the Libor-fixing process. According to Ellis \cite{Ellis}, studies carried out so far on Libor manipulation have not analyzed the situation thoroughly. In fact, proof of financial misconduct was based on internal bank e-mails and the confessions of bank officials.

The aim of this paper is to fill some gaps in the financial literature in several areas. First, we introduce a novel and powerful tool for analyzing economic time series: the Complexity Entropy Causality Plane (CECP). One of the key features of our proposed method is that it enables the discrimination of various stochastic and chaotic regimes. It therefore goes beyond traditional econometric analysis and could highlight changes in the hidden structure of complex systems such as the financial market. Second, we offer an exhaustive analysis of different interest rates in the United Kingdom (UK) market.  Third, our results back the investigations made by financial surveillance authorities such as the US Department of Justice and the UK Financial Services Authority regarding Libor settings.

Unlike previous studies, which analyzed interest rate spreads, we study the raw Libor time series. Our methodology has two advantages: (i) it makes no assumption about the underlying stochastic process governing interest rates, and (ii) by working with one series at a time, it avoids the problem caused by the interaction of (possibly) different stochastic dynamics.

This paper is organized as follows. In section \ref{sec:literature} we present a brief review of previous analyses of the Libor evolution rate. In section \ref{sec:itq} we describe the method based on Information Theory. In section \ref{sec:data} we describe the data used in the paper. In section \ref{sec:results} we present our results and in section \ref{sec:conclusions} we draw our main conclusions. 

\section{Brief literature review \label{sec:literature}}

\subsection{The efficient market hypothesis}
Since the seminal work of Bachelier \cite{Bachel}, prices in a competitive market have been modeled as a memoryless stochastic process. This is because, according to the Efficient Market Hypothesis, prices fully reflect all available information \cite{Fama76}. Therefore, interest rates should also follow approximately a random walk \cite{Fama1975Interest}. If the Libor was manipulated, a deterministic process will contaminate the pure stochastic process that governs interest rates in a competitive market. This situation will produce a deterioration in the informational efficiency of the time series. It is known that informational efficiency can vary over time. This may be due to several reasons. For example, Bariviera \cite{Bariviera11} argued that informational efficiency in the Thai stock market is influenced by liquidity constraints, while  Bariviera et al. \cite{BaGuMa12} detailed the influence of the 2008 financial crisis on the memory endowment of the European fixed income market. Consequently we aim to detect, by means of permutation information theory quantifiers, if this deterioration is contemporary of the manipulation that was detected by market authorities. 

\subsection{The Libor case}
The British Bankers Association (BBA) defines Libor as ``...the rate at which an individual Contributor Panel bank could borrow funds, were it to do so by asking for and then accepting inter-bank offers in reasonable market size, just prior to 11:00 [a.m.] London time''. 

Every London business day, each bank in the Contributor Panel (selected banks from BBA) makes a blind submission such that each banker does not know the quotes of the other bankers. A compiler (Thomson Reuters) then averages the second and third quartiles. This average is published and represents the Libor rate on a given day. In other words, Libor is a trimmed average of the expected borrowing rates of leading banks. Libor rates are published for several maturities and currencies.

The effect on the economy of underhand Libor rigging is deep and wide. On the one hand, it provides incorrect information on the true interbank lending cost and introduces an incorrect signal for Libor-linked contracts. On the other hand and, as highlighted by Stenfors \cite{Stenfors}, probably more importantly, it corrupts a ``key variable in the first stage of the monetary transmission mechanism''.

Many debt instruments issued at variable rate link the interest coupons to Libor. This variable rate is often defined as ``Libor$+\Delta$'', where ``$\Delta$'' is a premium that is usually linked to borrower's risk. If Libor underestimates the true lending cost (the intrinsic time value of money), borrowers gain at the expense of lenders.  A lower Libor induces lower mortgage rates, which makes it easier to buy homes and increase the volume of the real estate market at the expense of other goods markets. This artificially inflates home prices and related goods such as furniture and construction services. More generally, wrong interest rates modify consumption time preferences. Consequently, it is a topic worth researching and scrutinizing.

Despite its important economic consequences, this topic has been overlooked by mainstream economics and most research has been done by journalists.
The importance of a good pricing system is based on its usefulness for making decisions. As Hayek \cite{Hayek45} affirmed, ``we must look at the price system as such a mechanism for communicating information if we want to understand its real function''. If the price system is contaminated but perceived as fair, the effect could also reach the real economy and make it difficult to find a way out of the financial crisis. 

Taylor and Williams \cite{TaylorWilliams2009} documented the detachment of the Libor rate from other market rates such as Overnight Interest Swap (OIS), Effective Federal Fund (EFF),  Certificate of Deposits (CDs), Credit Default Swaps (CDS), and Repo rates.
Snider and Youle \cite{SniderYoule} studied individual quotes in the Libor bank panel and  found that Libor quotes in the US were not strongly related to other bank borrowing cost proxies.

Abrantes-Metz et al. \cite{AbrantesMetz2011} analyzed the distribution of the Second Digits (SDs) of daily Libor rates between 1987 and 2008 and compared it with uniform and Benford's distributions. If we take into account the whole period, the null hypothesis that the empirical distribution follows either the uniform or Benford's distribution cannot be rejected. However, if we take into account only the period after the subprime crisis, the null hypothesis is rejected. This result calls into question the ``aseptic'' setting of Libor.
Monticini and Thornton \cite{Monticini20133} found evidence of Libor underreporting after analyzing the spread between 1-month and 3-month Libor and the rate of Certificate of Deposits using the Bai and Perron \cite{BaiPerron1998} test for multiple structural breaks. 

The economic consequences of a ``contaminated'' Libor can be diverse and can have an uneven impact that affects several economic actors.
The lack of integrity of Libor as an information signal gives market participants a wrong proxy of borrowing costs, thus providing a bad rate for pricing financial products. 

With regard to the fixed income market, emerging markets such as Argentina, Chile and Colombia issue sovereign bonds that are linked to Libor 
(see Terce\~{n}o and Guercio \cite{TercenoGuercio2011}). 
Downward misreporting affects fixed income valuation in the secondary market, which leads to an overvaluation of prices and a reduction in the internal rate of return of bonds. This increases fixed income portfolios and leads to better public accounts in those countries (due to the lower interest load) and lower returns of fixed income mutual funds. 

Finally, and unlike medium- and long-term markets (corporate bonds and stocks), the money market plays a key role as a provider of liquidity to the financial sector. Interest rates disclosed by the market are therefore used as a measure of market liquidity and credit risk. Wrong market signals could lead to improper monetary policies by the central banks.

\section{Symbolic time series analysis \label{sec:itq}}

\subsection{The Bandt and Pompe method}

Many real world phenomena, such as the one under analysis in this paper, require the detailed analysis of observations at different time positions. One of the goals of time series analysis is to describe the nature of the generating process. Consequently, we can assume that a natural starting point for this task is to find the appropriate probability density function (PDF) associated to the time series. Among the several alternatives for PDF estimation, the Bandt and Pompe (BP) \cite{BandtPompe02} methodology has the advantage of considering time causality in its estimation. This symbolic methodology is robust to the presence of (observational) noise and requires no \textit{a priori} model assumption. The starting point of this method is to consider the ordinal structure of $D-$dimensional partitions of the time series. ``Partitions'' are devised  by comparing the order of neighboring relative values rather than by apportioning amplitudes according to different levels.

Given a time series ${\mathcal S}(t) = \{ x_t ; t = 1, \cdots , N \}$,
an embedding dimension $D > 1$  $(D \in {\mathbb N}$,
and an embedding delay $\tau$ $(\tau \in {\mathbb N})$, the
BP-pattern of order $D$ generated by
\begin{equation}
\label{eq:vectores}
s \mapsto \left(x_{s-(D-1)\tau},x_{s-(D-2)\tau},\cdots,x_{s-\tau},x_{s}\right) \ ,
\end{equation}
is the one to be considered.
To each time $s$, Bandt and Pompe method (BP) assigns a $D$-dimensional vector that results from the evaluation of the
time series at times $s - (D - 1) \tau, s-(D-2)\tau, \cdots , s - \tau, s$.
Clearly, the higher value of $D$, the more information about ``the past'' is incorporated
into the ensuing vectors.
By the ordinal pattern of order $D$ related to the time $s$, BP mean the permutation
$\pi = (r_0, r_1, \cdots , r_{D-1})$ of $(0, 1, \cdots ,D - 1)$ defined by
$x_{s-r_{D-1} \tau} \le  x_{s-r_{D-2} \tau} \le \cdots \le x_{s-r_{1} \tau}\le x_{s-r_0 \tau}$. 
In this way the vector defined by Eq. (\ref{eq:vectores}) is converted into a definite symbol $\pi$.
So as to get a unique result BP consider that $r_i < r_{i-1}$ if $x_{s-r_{i} \tau} = x_{s-r_{i-1} \tau}$.
This is justified if the values of ${x_t}$ have a continuous distribution so that equal values are
very unusual.

For all the $D!$ possible orderings (permutations)
$\pi_i$ when  embedding dimension is $D$, their associated relative
frequencies can be naturally computed according to the number of
times this particular order sequence is found in the time series,
divided by the total number of sequences,
\begin{equation}
\label{eq:frequ}
p(\pi_i)= \frac{\sharp \{s|s\leq N-(D-1)\tau ; (s) \quad \texttt{has type}~\pi_i \}}{N-(D-1)\tau} \ .
\end{equation}
In the last expression the symbol $\sharp$ stands for ``number".
Thus, an ordinal pattern probability distribution $P = \{ p(\pi_i), i = 1, \cdots, D! \}$
is obtained from the time series.

As mentioned above, the ordinal-pattern's associated PDF is invariant with respect to nonlinear monotonous
transformations. Accordingly, nonlinear drifts or scalings artificially introduced by a
measurement device will not modify the quantifiers' estimation, a
nice property if one deals with experimental data (see e.g. Saco et al. \cite{Saco2010}).
These  advantages make the BP approach more
convenient than conventional methods based on range partitioning.
Additional advantages of the  method reside in
{\it i)\/} its simplicity (we need  few parameters: the pattern length/embedding
dimension $D$ and the embedding delay $\tau$) and
{\it ii)\/} the extremely fast nature of the pertinent calculation-process \cite{Keller2005}.
The BP methodology can be applied not only  to time series representative
of low dimensional dynamical systems but also to any type of time
series (regular, chaotic, noisy, or reality based).
In fact, the existence of an attractor in the $D$-dimensional phase space in not assumed.
The only condition for the applicability of the BP method is  a very
weak stationary assumption: for $k=D$, the probability
for $x_t < x_{t+k}$ should not depend on $t$.

For a review of BP's methodology and its applications to physics, biomedical and econophysics signals
see Zanin et al. \cite{Zanin2012}.

\subsection{Permutation Information Theory quantifiers}

We introduce an approach that uses quantifiers derived from Information Theory: entropy and statistical complexity. In order to evaluate these quantifiers, we use the Bandt and Pompe permutation method \cite{BandtPompe02} to estimate the probability distribution associated with the time series. As is widely known, symbolic time-series analysis methods that discretize the raw time series into a corresponding sequence of symbols are able to analyze nonlinear data efficiently with low sensitivity to noise. However, it is not easy to find a meaningful symbolic representation of the original series. 
The Bandt and Pompe approach is the only symbolization technique of those in popular use that takes into account the time causality of the system's dynamics. Important details concerning the ordinal structure of the time series are then revealed.

We calculate two magnitudes: the permutation entropy and the permutation statistical complexity of each series.
According Shannon and Weaver \cite{book:shannon1949}, given a discrete probability distribution 
$P=\{ p_i \in {\mathbb R} ; p_i \geq 0 ; i= 1, \cdots, M\}$, with 
$\sum_{i=1}^M p_i = 1$,
Shannon entropy is defined as:
\begin{equation}
{\cal S}[P]= -\sum_{i=1}^M p_i \ln{p_i}.
\label{eq:entropy}
\end{equation}
 
This quantifier equals zero if the patterns are fully deterministic and reaches its maximum value for a uniform distribution. 
Using informational entropy to study economic phenomena is not new. In fact, Theil and Leenders\cite{TheilLeenders65}, Fama \cite{Fama65entropy}, and Dryden \cite{Dryden68} may be considered seminal papers in this field. More recently,  Martina et al. \cite{Martina2011} and Ortiz et al. \cite{OrtizCruz12} applied entropy and multiscale entropy analysis to assess crude oil price efficiency. Alvarez-Ram\'irez et al.\cite{AlvarezRamirez2012} also used entropy methods to quantify the dynamics of the informational efficiency of the US stock market over the last 70 years. Iglesias and de Almeida \cite{IglesiasDeAlmeida2012} showed that some market models lead to a condensed state and, consequently,  to a state of minimum entropy. In order to recover a Boltzmann-Gibbs distribution, some regulations must be included in the market. This study highlighted the importance of loans in order to achieve an state of maximum entropy at a given market.

However, analyzing time series by means of Shannon entropy alone could be insufficient. As stressed in Feldman and Crutchfield \cite{FeldmanCrutchfield98} and Feldman et al. \cite{FeldmanMcTague08}, an entropy measure does not quantify the degree of structure or patterns present in a process and a measure of statistical complexity must be introduced into the analysis in order to characterize a system's organizational properties. 
A family of statistical complexity measures, based on the functional form developed by L\'opez-Ruiz et al. \cite{LMC95}, is defined in Mart\'in et al. \cite{Martin2003} and Lamberti et al. \cite{Lamberti2004119} as:
\begin{equation}
{\cal C}_{JS}= {\cal Q}_J[P,P_e] {\cal H}_S [P]
\label{eq:complexity}
\end{equation}
where ${\cal H}_S [P]=S[P]/S_{\max}$ is the normalized Shannon entropy, $P$ is the discrete probability distribution associated with the time series under analysis, $P_e$ is the uniform distribution and ${\cal Q}_J [P,P_e]$ is the so-called disequilibrium:
$${\mathcal Q}_J [P,P_e] = Q_0 \{ S[ (P+P_e)/2 ] - S[P]/2 - S[P_e]/2 \}$$
with $Q_0$ a normalization constant.

This disequilibrium is defined in terms of the Jensen-Shannon divergence, which quantifies the difference between two probability distributions. Mart\'in et al. \cite{paper:martin2006} demonstrates the existence of upper and lower bounds for generalized statistical complexity measures such as ${\cal C}_{JS}$. Additionally, as highlighted in Soriano et al. \cite{Soriano2011a}, the permutation complexity is not a trivial function of the permutation entropy because it is based on two probability distributions. 

The planar representation of these quantifiers is introduced in efficiency analysis and has been successfully used to: rank efficiency in stock markets \cite{Zunino2010a,ZuninoCausality10}; rank efficiency in commodity markets \cite{ZuninoPermutation11}; link informational efficiency with sovereign bond ratings \cite{Zunino2012}; and assess the impact of the establishment of a common currency and a deep and wide financial crisis in  European sovereign bonds time series \cite{Bariviera2013epjb}.

\section{Data}
\label{sec:data}

Following Taylor and Williams \cite{TaylorWilliams2009}, we selected several UK benchmark interest rates. In addition to Libor we considered other relevant interest rates in the UK market: the Repo benchmark (REPO), Overnight Interest Swaps (OIS) and the Sterling OverNight Index Average (SONIA).
The data span was from 17/05/1999 to 08/09/2014, with a total of 3996 data points, except for the UK OIS rate, which began on 19/01/2004 (2776 data points). All data were retrieved from DataStream.
One main difference among these rates is that Libor and Repo are calculated as the arithmetic averages of the middle two quartiles of the estimations delivered by selected banks, whereas the other two rates represent the averages of the actual transactions.

\section{Results}
\label{sec:results}

Using the above data and the methodology described in Section \ref{sec:itq}, we computed the permutation entropy and permutation statistical complexity for each series using sliding windows. The rationale behind the use of a moving sample was that it enable us to study the evolution of these quantifiers during the period under examination. Our sliding windows contained $N=300$ data points, the frequency was daily ($\tau=1$), and the step between each window was $\delta=20$ data points. In this way we obtained $184$ estimation periods. 
The dates for the beginning and end of each period are provided in the appendix. 

It is worth mentioning that when we performed our analysis for pattern length (embedding dimension) $D=\{3,4,5\}$ using windows sample lengths $300 \leq N \leq 1500$, we obtained similar results. 
We believe that with a pattern length $D=4$ and 300 data points we are able to capture the dynamics of the series under analysis. Each window spans approximately one year and moves one month ahead. In this way changes in the underlying stochastic process are taken into account.

Following Zunino et al. \cite{Zunino2010a,ZuninoCausality10,ZuninoPermutation11} and Rosso et al. \cite{RossoNoise07}, we use the Bandt and Pompe
\cite{BandtPompe02} permutation method to obtain the appropriate probability distribution function. 
If a series is purely random, permutation entropy is maximized and permutation statistical complexity is minimized. Since we work with normalized quantifiers, the maximum efficiency point of the Complexity Entropy Causality Plane (CECP) is $(1,0)$. We argue that, without manipulation, the location of quantifiers across windows should be stable or at least follow no predictable path.

\subsection{Preliminary analysis}

\begin{figure}[!ht]
\center
\includegraphics[scale=.45]{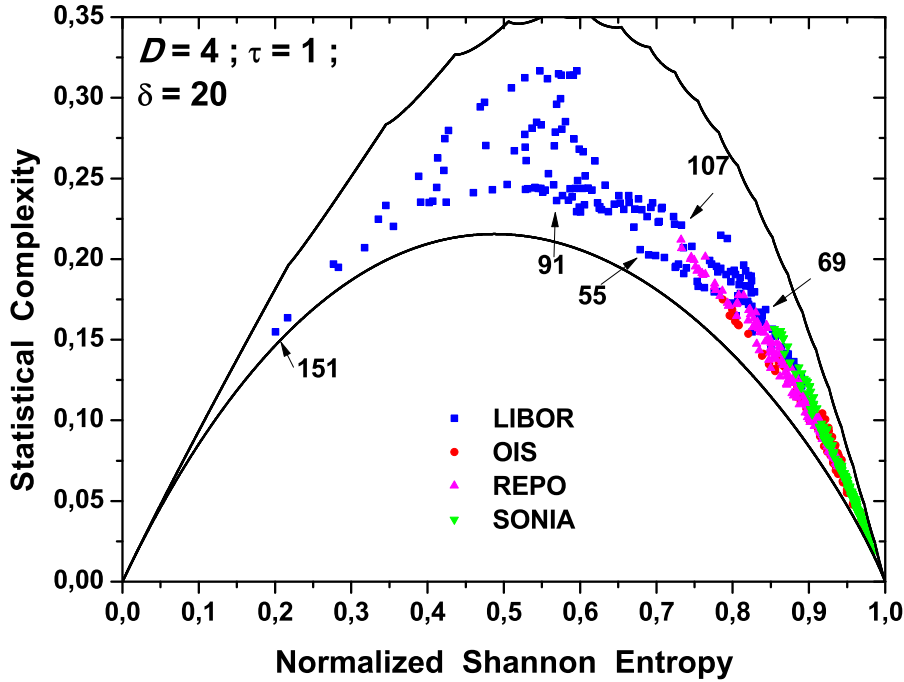}
\caption{
Complexity Entropy Causality Plane for $D=4$, $\tau=1$, $\delta=20$ of UK interest rates:
the London Interbank Offered Rate (LIBOR), the Repo benchmark (REPO), Overnight Interest Swaps (OIS) and Sterling OverNight Index Average (SONIA).
The data span was from 17/05/1999 until  08/09/2014,  except for UK OIS rate, which began on 19/01/2004.
The solid lines represent the upper and lower bounds of the quantifiers as computed by Mart\'in et al. \cite{paper:martin2006}
}
\label{fig:CECP_UK}
\end{figure}

In Figure \ref{fig:CECP_UK} we show the Complexity Entropy Causality Plane of UK data. Each point reflects the calculations of permutation entropy and permutation complexity for a period of the sliding window. We can see that SONIA, REPO and OIS rates  occupy the bottom right-hand area of the CECP. This area may be associated with random processes that exhibit little or no memory (this behavior is compatible with colored noises). In other words, information about past prices is (almost) fully embedded in the current price, which according to Fama \cite{Fama70}, means that the market is informationally efficient in its weak form. In contrast, the Libor rate is scattered throughout much of the CECP. 

A simple way of observing the difference in the informational content of each time series is by means of the descriptive statistics of the permutation entropy estimations (see Table \ref{tab:statistics}). Libor  has a mean permutation entropy of approximately 0.67, whereas ${\cal H}_{OIS}^{mean} \approx 0.91$ ${\cal H}_{SONIA}^{mean} \approx 0.94$ and ${\cal H}_{Repo}^{mean} \approx 0.87$. We preform the test of equality of means, in order to assess if Libor mean permutation entropy is equal to the permutation entropy of each of the other series. As shown in Table 3 of the appendix, the null hypothesis is rejected. This result confirm the different path behavior of the series under analysis. We would like to highlight that the range of the permutation entropy estimations differs enormously, and that the minimum value for OIS, SONIA and Repo is around 0.75, whereas the minimum for Libor is 0.20.

\begin{table}[htbp]
  \centering
  \caption{Descriptive statistics of the permutation entropy of each series, and test of mean equality of each series with respect to Libor.}
    \begin{tabular}{rrrrr}
    \toprule
          & \multicolumn{1}{c}{Libor} & \multicolumn{1}{c}{OIS} & \multicolumn{1}{c}{Repo} & \multicolumn{1}{c}{SONIA} \\
    \midrule
    Mean  & 0.66915 & 0.91410 & 0.86932 & 0.94137 \\
    Median & 0.68425 & 0.92786 & 0.88005 & 0.94930 \\
    Std. Dev & 0.15697 & 0.04227 & 0.05302 & 0.03020 \\
    Min   & 0.20053 & 0.78696 & 0.73237 & 0.85500 \\
    Max   & 0.90555 & 0.97988 & 0.94743 & 0.98502 \\
    F     &       & 13.78748 & 8.76667 & 27.01272 \\
    p-value &       & 0.00000 & 0.00000 & 0.00000 \\
    \bottomrule
    \end{tabular}%
  \label{tab:statistics}%
\end{table}%

\begin{figure}[!ht]
 \center
   \includegraphics[scale=.3]{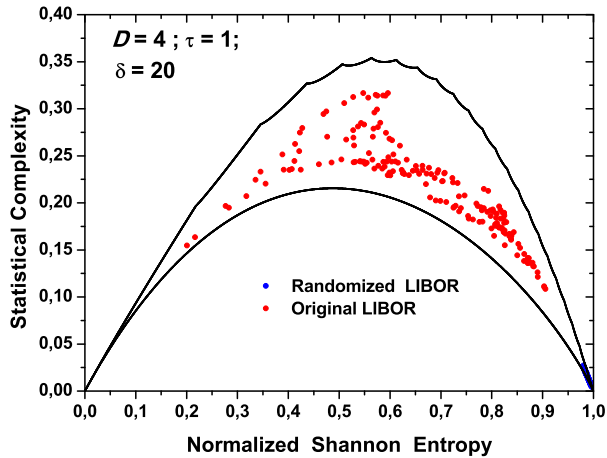}
   \caption{Complexity Entropy Causality Plane of original Libor series and randomized Libor series.}
   \label{fig:CECP_UK_RANDOM}
\end{figure}

In order to highlight that the planar location of the quantifiers were not obtained by chance, we display in Figure \ref{fig:CECP_UK_RANDOM} the results of the original Libor series and the those of the randomized Libor series. When we shuffle data and thus, destroy all non trivial correlations, permutation informations quantifiers move close to the $(1,0)$ corner. This picture shows that the proposed quantifiers capture the hidden correlation structure of data.

\subsection{Temporal evolution of the informational efficiency}

\begin{figure}[!ht]
 \center
   \includegraphics[scale=0.45]{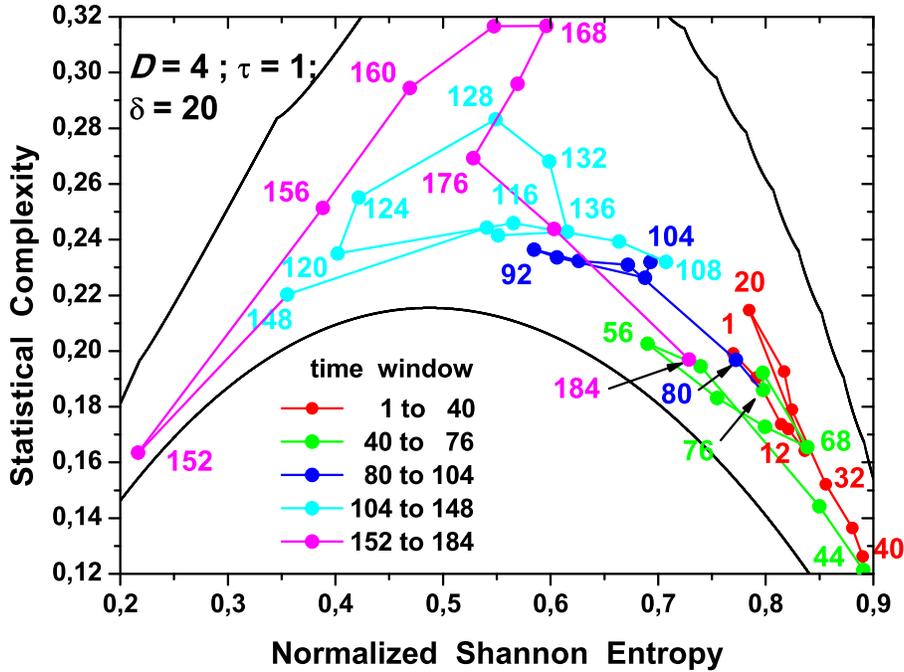}
   \caption{Complexity Entropy Causality Plane for $D=4$, $\tau=1$, $\delta=20$. Conceptual scheme of the movement of the permutation information quantifiers for the  evolution of UK Libor.}
   \label{fig:CECP_UK_esquema}
\end{figure}

We can use figure 1 to analyze the direction of movement over time, focusing on the Libor rate (see Fig \ref{fig:CECP_UK_esquema}). 

We can clearly see that the Libor rate performed relatively well (i.e. following dynamics that are compatible with a random walk) until the estimation of time window 69, $( H, C )^{(LIBOR)} = (0.843,~0.169)$, which corresponds to the period between August 2004 and September 2005, where there is a trend left and upwards. 
This trend continues until time window 91, $( H, C )^{(LIBOR)} = (0.569,~0.236)$, which corresponds to the period between April 2006 and June 2007, where we see a backward movement in the path until window 107, $( H , C)^{(LIBOR)} = ( 0.732,~0.221)$, corresponding to the period between July 2007 and August 2008. This point reflects several distinctive characteristics. 

First, we clearly observe a path toward areas of inefficiency, i.e. higher permutation complexity and lower permutation entropy. Second, there is a clear separation between Libor and the other interest rates in the CECP. Third, this period coincides with the beginning of the upheaval in the financial markets due to the subprime crisis. And fourth, Mollenkamp and Whitehouse \cite{MollenkampWhitehouse} published their seminal article on possible Libor manipulation in The Wall Street Journal. According to \cite{Reuters2012}, the Federal Reserve Bank of New York was aware of manipulative activities as early as 2008. 

The temporal movements of the points in the CECP cannot be attributed to chance but to either a change in the underlying statistical laws governing the money market or to hidden manipulation by market participants. Although statistical methods themselves cannot prove the existence of manipulation, they could be useful early warning devices for surveillance authorities such as central banks or exchange commissions to propose in-depth investigations.

Based on Figure \ref{fig:CECP_UK}, we drew a conceptual scheme of the movement of the permutation information quantifiers (see Figure \ref{fig:CECP_UK_esquema}). 
We took a subsample of the sliding windows in a proportion of 1:4 in order to avoid a saturated figure and linked four consecutive windows where the numbers indicate the direction of the movement.  

The path toward a more informationally inefficient area continues until window 151, 
$( H, C )^{(LIBOR)} = ( 0.201,~0.155)$, which corresponds to the period between November 2010 and January 2012 (see Fig. \ref{fig:CECP_UK}). 
After this period, a U-turn is observed and the trend is reversed, as we can see in Figure \ref{fig:CECP_UK_esquema}. 
Our estimations reflect a slow but constant improvement in the informational efficiency of the Libor time series. 
In Figure \ref{fig:CECP_UK}, the point corresponding to window 183, 
$( H, C)^{(LIBOR)} = ( 0.670,~0.074)$, corresponding to the period between April 2013 and June 2014, is located at a similar entropy level to that of window 55, $( H, C)^{(LIBOR)} = (0.678,~0.127)$, which corresponds to the period between July 2003 and August 2004. This improvement appears to be consistent with the revelation of the Libor scandal and the subsequent reform in Libor setting characteristics. 

This is particularly important because the inefficiency path has not been followed by other interest rates. If Libor behavior is clearly different from that of other interest rate behavior, what is the difference between Libor and the other selected interest rates? The answer could lie in the highly intrinsic characteristics of the Libor rate. In fact, Libor and Repo methodology on the one hand and OIS and SONIA methodology on the other differ substantially. Whereas OIS and SONIA both reflect an average of actual transactions, Libor and Repo benchmark rates reflect trimmed averages of the self-estimated borrowing costs of a selected group of banks. Libor and Repo benchmarks do not necessarily reflect effective market transaction. In fact, the Libor quotes of individual banks are non-binding. We could say that Libor reflects the perceptions or desires of decision-makers (to the best of their knowledge) better than a bank's true borrowing costs. Libor therefore does not reflect an equilibrium price like a Walrasian auctioneer does. More closely, it reflects the rate at which panel banks would like to borrow from their colleagues.

We would like to reiterate that the purpose of our methodology was not to detect data manipulation. However, by using this methodology we were able to detect a significant change in the stochastic process governing the Libor time series that overlaps with the landmark moments of the beginning of Libor manipulation and the detection of the Libor scandal. Bearing in mind that several banks admitted their participation in manipulation activities and paid enormous fines for abusing their market positions (see e.g. \cite{Liborthenandnow} and \cite{FTemail}), we can affirm that our methodology, whose purpose is to describe the stochastic and/or chaotic features of time series, is a suitable tool for detecting exogenous forces that alter normal price setting in a competitive market. 

\begin{figure}[!ht]
 \center
   \includegraphics[scale=.45]{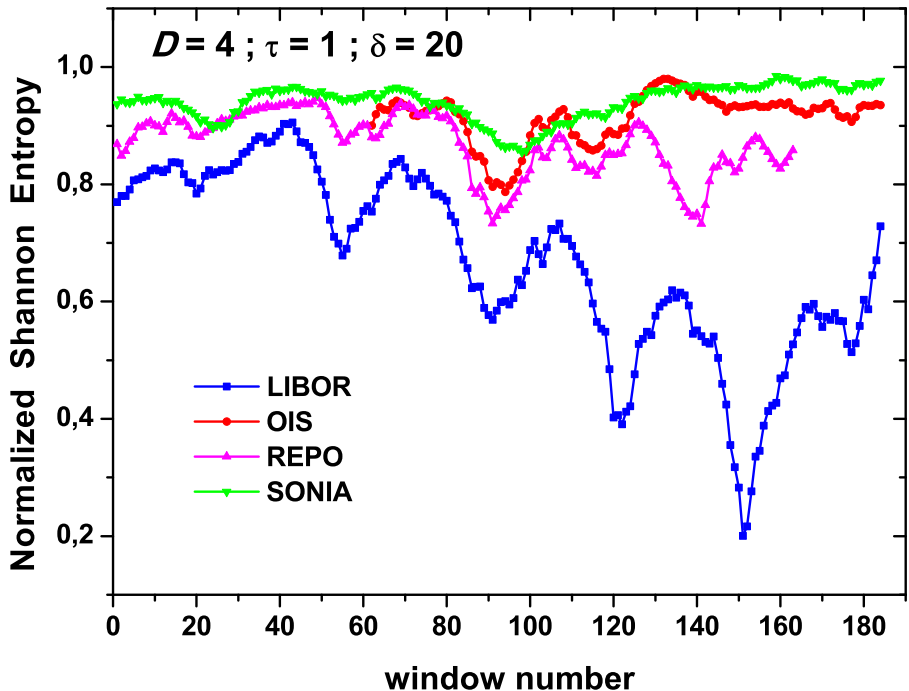}
   \caption{Permutation entropy evolution, computed for $D=4$, $\tau=1$, $\delta=20$,  of the various UK interest rates: 
the London Interbank Offered Rate (LIBOR), the Repo benchmark (REPO), Overnight Interest Swaps (OIS) and the Sterling OverNight Index Average (SONIA). The data span was from 17/05/1999 to 08/09/2014, except for the UK OIS rate, which began on 19/01/2004.}
   \label{fig:entro_evol_UK}
\end{figure}

To clarify our claim, Figure \ref{fig:entro_evol_UK} shows the temporal evolution of the permutation entropy of all the interest rates. We can clearly see that the permutation entropies of the rates that reflect actual transactions (OIS and SONIA) are relatively high and that their movement is bounded between 0.8 and 1. On the other hand, the permutation entropy of the Libor rate deteriorates progressively, reaching an absolute minimum ($H_S\approx 0.2$) in period 151. After this point there is a steady recovery in the higher levels of permutation entropy. If we set an imaginary linear trend line, we find that although Libor recovered a path towards more random dynamics, the trend during the whole period was to decrease. Libor is the only interest rate that showed this trend.

The behavior of Repo is probably half-way between that of Libor and that of the other rates. Although the method used to calculate Repo is similar to that used to calculate Libor, Repo assumes an underlying asset as collateral. This could make Repo less prone to manipulation. 

\section{Conclusions}
\label{sec:conclusions}

We have applied a novel technique in Finance to detect changes in the underlying stochastic/chaotic process that governs the movements of interest rates. This technique uses two concepts borrowed from Information Theory: permutation entropy and permutation statistical complexity. Using sliding windows we reveal the mutating stochastic character of the interest rates time series. 

Unlike previous works we provide a quantitative methodology for accessing Libor dynamics.
Antitrust law enforcement is complex  because
manipulation and fraud can be elegantly camouflaged. It is difficult for a
statistical procedure to be accepted as legal proof in
a court of law. However, its use by surveillance authorities makes
any attempt at manipulation more costly and more difficult to
continue. We therefore view our proposal as a market watch
mechanism that could make attempts at manipulation and/or collusion 
more difficult in the future. Moreover, an efficient
overseeing mechanism could increase the incentive to apply for
leniency at earlier stages of manipulation
\cite{AbrantesSokol2012}.

Libor clearly moved far away from a memoryless stochastic process. The low permutation entropy and high statistical complexity exhibited in the peak of the financial crisis clearly indicates the existence of a complex structure of temporal correlations. 

Symbolic analysis reveals the peculiar behavior of Libor \textit{vis-\`a-vis} other selected rates. We cannot affirm that manipulation was the cause of this strange behavior. However, given the outcome of the Libor investigations conducted by several financial authorities, we could say that our findings are consistent with some kind of ``non-invisible hand''. We believe that surveillance authorities should use advanced statistical techniques such as the one proposed in this paper to enhance market discipline. 

We would like to stress that our methodology is easy to implement. Also, as we have shown in this paper, updating data every month provides an overview of how market prices are behaving.

\section*{Acknowledgements}
MBG, LBM and OAR are members of the National Research Career of CONICET Argentina.  
MBG, LBM and OAR  acknowledge support by CONICET, Argentina.
 
We would like to thank the anonymous reviewer and the editor who evaluated this manuscript for their thoughtful comments and suggestions.

Author contribution: All authors contributed equally to the paper.

\bibliographystyle{plain}  
\bibliography{liborbib}

\appendix{Appendix}

\section{Dates of each estimation period }
Table \ref{tab:periods} and Table \ref{tab:periods2} contain the initial and final dates of each sliding window.
\footnotesize

\begin{table*}[htbp]
  \centering
  \caption{Initial and final dates of each estimation period}
    \begin{tabular}{>{\tiny}r >{\tiny}r >{\tiny}r |>{\tiny}r >{\tiny}r >{\tiny}r  |>{\tiny}r >{\tiny}r >{\tiny}r }
    \multicolumn{1}{c}{\#} & \multicolumn{1}{c}{Begins} & \multicolumn{1}{c}{Ends} & \multicolumn{1}{c}{\#} & \multicolumn{1}{c}{Begins} & \multicolumn{1}{c}{Ends} & \multicolumn{1}{c}{\#} & \multicolumn{1}{c}{Begins} & \multicolumn{1}{c}{Ends} \\
    \toprule
    1     & 17-May-99 & 7-Jul-00 & 31    & 3-Sep-01 & 25-Oct-02 & 61    & 22-Dec-03 & 11-Feb-05 \\    
    2     & 14-Jun-99 & 4-Aug-00 & 32    & 1-Oct-01 & 22-Nov-02 & 62    & 19-Jan-04 & 11-Mar-05 \\
    3     & 12-Jul-99 & 1-Sep-00 & 33    & 29-Oct-01 & 20-Dec-02 & 63    & 16-Feb-04 & 8-Apr-05 \\
    4     & 9-Aug-99 & 29-Sep-00 & 34    & 26-Nov-01 & 17-Jan-03 & 64    & 15-Mar-04 & 6-May-05 \\
    5     & 6-Sep-99 & 27-Oct-00 & 35    & 24-Dec-01 & 14-Feb-03 & 65    & 12-Apr-04 & 3-Jun-05 \\
    6     & 4-Oct-99 & 24-Nov-00 & 36    & 21-Jan-02 & 14-Mar-03 & 66    & 10-May-04 & 1-Jul-05 \\
    7     & 1-Nov-99 & 22-Dec-00 & 37    & 18-Feb-02 & 11-Apr-03 & 67    & 7-Jun-04 & 29-Jul-05 \\
    8     & 29-Nov-99 & 19-Jan-01 & 38    & 18-Mar-02 & 9-May-03 & 68    & 5-Jul-04 & 26-Aug-05 \\
    9     & 27-Dec-99 & 16-Feb-01 & 39    & 15-Apr-02 & 6-Jun-03 & 69    & 2-Aug-04 & 23-Sep-05 \\
    10    & 24-Jan-00 & 16-Mar-01 & 40    & 13-May-02 & 4-Jul-03 & 70    & 30-Aug-04 & 21-Oct-05 \\
    11    & 21-Feb-00 & 13-Apr-01 & 41    & 10-Jun-02 & 1-Aug-03 & 71    & 27-Sep-04 & 18-Nov-05 \\
    12    & 20-Mar-00 & 11-May-01 & 42    & 8-Jul-02 & 29-Aug-03 & 72    & 25-Oct-04 & 16-Dec-05 \\
    13    & 17-Apr-00 & 8-Jun-01 & 43    & 5-Aug-02 & 26-Sep-03 & 73    & 22-Nov-04 & 13-Jan-06 \\
    14    & 15-May-00 & 6-Jul-01 & 44    & 2-Sep-02 & 24-Oct-03 & 74    & 20-Dec-04 & 10-Feb-06 \\
    15    & 12-Jun-00 & 3-Aug-01 & 45    & 30-Sep-02 & 21-Nov-03 & 75    & 17-Jan-05 & 10-Mar-06 \\
    16    & 10-Jul-00 & 31-Aug-01 & 46    & 28-Oct-02 & 19-Dec-03 & 76    & 14-Feb-05 & 7-Apr-06 \\
    17    & 7-Aug-00 & 28-Sep-01 & 47    & 25-Nov-02 & 16-Jan-04 & 77    & 14-Mar-05 & 5-May-06 \\
    18    & 4-Sep-00 & 26-Oct-01 & 48    & 23-Dec-02 & 13-Feb-04 & 78    & 11-Apr-05 & 2-Jun-06 \\
    19    & 2-Oct-00 & 23-Nov-01 & 49    & 20-Jan-03 & 12-Mar-04 & 79    & 9-May-05 & 30-Jun-06 \\
    20    & 30-Oct-00 & 21-Dec-01 & 50    & 17-Feb-03 & 9-Apr-04 & 80    & 6-Jun-05 & 28-Jul-06 \\
    21    & 27-Nov-00 & 18-Jan-02 & 51    & 17-Mar-03 & 7-May-04 & 81    & 4-Jul-05 & 25-Aug-06 \\
    22    & 25-Dec-00 & 15-Feb-02 & 52    & 14-Apr-03 & 4-Jun-04 & 82    & 1-Aug-05 & 22-Sep-06 \\
    23    & 22-Jan-01 & 15-Mar-02 & 53    & 12-May-03 & 2-Jul-04 & 83    & 29-Aug-05 & 20-Oct-06 \\
    24    & 19-Feb-01 & 12-Apr-02 & 54    & 9-Jun-03 & 30-Jul-04 & 84    & 26-Sep-05 & 17-Nov-06 \\
    25    & 19-Mar-01 & 10-May-02 & 55    & 7-Jul-03 & 27-Aug-04 & 85    & 24-Oct-05 & 15-Dec-06 \\
    26    & 16-Apr-01 & 7-Jun-02 & 56    & 4-Aug-03 & 24-Sep-04 & 86    & 21-Nov-05 & 12-Jan-07 \\
    27    & 14-May-01 & 5-Jul-02 & 57    & 1-Sep-03 & 22-Oct-04 & 87    & 19-Dec-05 & 9-Feb-07 \\
    28    & 11-Jun-01 & 2-Aug-02 & 58    & 29-Sep-03 & 19-Nov-04 & 88    & 16-Jan-06 & 9-Mar-07 \\
    29    & 9-Jul-01 & 30-Aug-02 & 59    & 27-Oct-03 & 17-Dec-04 & 89    & 13-Feb-06 & 6-Apr-07 \\
    30    & 6-Aug-01 & 27-Sep-02 & 60    & 24-Nov-03 & 14-Jan-05 & 90    & 13-Mar-06 & 4-May-07 \\
    \bottomrule
       \end{tabular}%
  \label{tab:periods}%
\end{table*}%

\newpage

\begin{table*}[htbp]
  \centering
  \caption{Initial and final dates of each estimation period (cont.)}
    \begin{tabular}{>{\tiny}r >{\tiny}r >{\tiny}r |>{\tiny}r >{\tiny}r >{\tiny}r  |>{\tiny}r >{\tiny}r >{\tiny}r }
    \multicolumn{1}{c}{\#} & \multicolumn{1}{c}{Begins} & \multicolumn{1}{c}{Ends} & \multicolumn{1}{c}{\#} & \multicolumn{1}{c}{Begins} & \multicolumn{1}{c}{Ends} & \multicolumn{1}{c}{\#} & \multicolumn{1}{c}{Begins} & \multicolumn{1}{c}{Ends}   \\ \toprule
    91    & 10-Apr-06 & 1-Jun-07 & 121   & 28-Jul-08 & 18-Sep-09 & 151   & 15-Nov-10 & 6-Jan-12 \\
    92    & 8-May-06 & 29-Jun-07 & 122   & 25-Aug-08 & 16-Oct-09 & 152   & 13-Dec-10 & 3-Feb-12 \\
    93    & 5-Jun-06 & 27-Jul-07 & 123   & 22-Sep-08 & 13-Nov-09 & 153   & 10-Jan-11 & 2-Mar-12 \\
    94    & 3-Jul-06 & 24-Aug-07 & 124   & 20-Oct-08 & 11-Dec-09 & 154   & 7-Feb-11 & 30-Mar-12 \\
    95    & 31-Jul-06 & 21-Sep-07 & 125   & 17-Nov-08 & 8-Jan-10 & 155   & 7-Mar-11 & 27-Apr-12 \\
    96    & 28-Aug-06 & 19-Oct-07 & 126   & 15-Dec-08 & 5-Feb-10 & 156   & 4-Apr-11 & 25-May-12 \\
    97    & 25-Sep-06 & 16-Nov-07 & 127   & 12-Jan-09 & 5-Mar-10 & 157   & 2-May-11 & 22-Jun-12 \\
    98    & 23-Oct-06 & 14-Dec-07 & 128   & 9-Feb-09 & 2-Apr-10 & 158   & 30-May-11 & 20-Jul-12 \\
    99    & 20-Nov-06 & 11-Jan-08 & 129   & 9-Mar-09 & 30-Apr-10 & 159   & 27-Jun-11 & 17-Aug-12 \\
    100   & 18-Dec-06 & 8-Feb-08 & 130   & 6-Apr-09 & 28-May-10 & 160   & 25-Jul-11 & 14-Sep-12 \\
    101   & 15-Jan-07 & 7-Mar-08 & 131   & 4-May-09 & 25-Jun-10 & 161   & 22-Aug-11 & 12-Oct-12 \\
    102   & 12-Feb-07 & 4-Apr-08 & 132   & 1-Jun-09 & 23-Jul-10 & 162   & 19-Sep-11 & 9-Nov-12 \\
    103   & 12-Mar-07 & 2-May-08 & 133   & 29-Jun-09 & 20-Aug-10 & 163   & 17-Oct-11 & 7-Dec-12 \\
    104   & 9-Apr-07 & 30-May-08 & 134   & 27-Jul-09 & 17-Sep-10 & 164   & 14-Nov-11 & 4-Jan-13 \\
    105   & 7-May-07 & 27-Jun-08 & 135   & 24-Aug-09 & 15-Oct-10 & 165   & 12-Dec-11 & 1-Feb-13 \\
    106   & 4-Jun-07 & 25-Jul-08 & 136   & 21-Sep-09 & 12-Nov-10 & 166   & 9-Jan-12 & 1-Mar-13 \\
    107   & 2-Jul-07 & 22-Aug-08 & 137   & 19-Oct-09 & 10-Dec-10 & 167   & 6-Feb-12 & 29-Mar-13 \\
    108   & 30-Jul-07 & 19-Sep-08 & 138   & 16-Nov-09 & 7-Jan-11 & 168   & 5-Mar-12 & 26-Apr-13 \\
    109   & 27-Aug-07 & 17-Oct-08 & 139   & 14-Dec-09 & 4-Feb-11 & 169   & 2-Apr-12 & 24-May-13 \\
    110   & 24-Sep-07 & 14-Nov-08 & 140   & 11-Jan-10 & 4-Mar-11 & 170   & 30-Apr-12 & 21-Jun-13 \\
    111   & 22-Oct-07 & 12-Dec-08 & 141   & 8-Feb-10 & 1-Apr-11 & 171   & 28-May-12 & 19-Jul-13 \\
    112   & 19-Nov-07 & 9-Jan-09 & 142   & 8-Mar-10 & 29-Apr-11 & 172   & 25-Jun-12 & 16-Aug-13 \\
    113   & 17-Dec-07 & 6-Feb-09 & 143   & 5-Apr-10 & 27-May-11 & 173   & 23-Jul-12 & 13-Sep-13 \\
    114   & 14-Jan-08 & 6-Mar-09 & 144   & 3-May-10 & 24-Jun-11 & 174   & 20-Aug-12 & 11-Oct-13 \\
    115   & 11-Feb-08 & 3-Apr-09 & 145   & 31-May-10 & 22-Jul-11 & 175   & 17-Sep-12 & 8-Nov-13 \\
    116   & 10-Mar-08 & 1-May-09 & 146   & 28-Jun-10 & 19-Aug-11 & 176   & 15-Oct-12 & 6-Dec-13 \\
    117   & 7-Apr-08 & 29-May-09 & 147   & 26-Jul-10 & 16-Sep-11 & 177   & 12-Nov-12 & 3-Jan-14 \\
    118   & 5-May-08 & 26-Jun-09 & 148   & 23-Aug-10 & 14-Oct-11 & 178   & 10-Dec-12 & 31-Jan-14 \\
    119   & 2-Jun-08 & 24-Jul-09 & 149   & 20-Sep-10 & 11-Nov-11 & 179   & 7-Jan-13 & 28-Feb-14 \\
    120   & 30-Jun-08 & 21-Aug-09 & 150   & 18-Oct-10 & 9-Dec-11 & 180   & 4-Feb-13 & 28-Mar-14 \\
          &       &       &       &       &       & 181   & 4-Mar-13 & 25-Apr-14 \\
          &       &       &       &       &       & 182   & 1-Apr-13 & 23-May-14 \\
          &       &       &       &       &       & 183   & 29-Apr-13 & 20-Jun-14 \\
          &       &       &       &       &       & 184   & 27-May-13 & 18-Jul-14 \\
    \bottomrule
       \end{tabular}%
  \label{tab:periods2}%
\end{table*}%

\end{document}